# Distributed Algorithm for Empty Vehicles Management in Personal Rapid Transit (PRT) network

Wiktor B. Daszczuk, Ph.D. - Institute of Computer Science, Warsaw University of Technology, Nowowiejska str. 15/19, 00-665 Warszawa, Poland, Tel. (+48)22-234-7812, Fax (+48)22-234-6091, Email: wbd@ii.pw.edu.pl,

Jerzy Mieścicki, Ph.D - Institute of Computer Science, Warsaw University of Technology, Nowowiejska str. 15/19, 00-665 Warszawa, Poland, Tel. (+48)22-234-7812, Fax (+48)22-234-6091, Email: jms@ii.pw.edu.pl,

Waldemar Grabski, M.Sc - Institute of Computer Science, Warsaw University of Technology, Nowowiejska str. 15/19, 00-665 Warszawa, Poland, Tel. (+48)22-234-7812, Fax (+48)22-234-6091, Email: wgr@ii.pw.edu.pl

**Abstract**: In this paper, an original heuristic algorithm of empty vehicles management in Personal Rapid Transit (PRT) network is presented. The algorithm is used for the delivery of empty vehicles for waiting passengers; for balancing the distribution of empty vehicles within the network; for providing an empty space for vehicles approaching a station, etc. Each of these tasks involves a decision on the trip which has to be done by a selected empty vehicle from its actual location to some determined destination. The decisions are based on a multi-parameter function, involving a set of factors and thresholds.

An important feature of the algorithm is that it does not use any central database of passenger input (demand) and locations of free vehicles. Instead, it is based on the local exchange of data between stations: on their states and on the vehicles they expect. Therefore, it seems well tailored for a distributed implementation.

The algorithm is uniform, meaning that the same basic procedure is used for multiple tasks using task-specific set of parameters.

## Introduction

Personal Rapid Transit (PRT) [1,2,3,4] is an urban transit system organized as a network, covering, for instance, a part of a city or a multi-terminal airport, a large exhibition area etc. The driverless (i.e. system-controlled) vehicles move along one-way tracks which are separated from a conventional traffic, e.g., through elevation of the tracks above the ground level. Typically, a vehicle can carry a group of 1 – 6 passengers. The term 'personal' means that a passenger, or a group of passengers, chooses the time as well as the destination of a trip freely. The system determines the best route for the trip, which is not necessarily the shortest one, and controls the vehicle movement during the voyage (acceleration/deceleration, preserving the separation between vehicles, passing intersections, avoiding traffic jams, etc.) . In addition to the control of these so-called *full* trips, the system manages the set of *empty* vehicles movement as well, which is analyzed in more detail below.

We can divide control algorithms of the PRT network into two levels:

- Coordination algorithms, used for the control of vehicles' movement following one another down the track, for the coordination at join-type intersections and for the control of the vehicle behavior inside stations and capacitors, as described in [5]. These coordination algorithms are not a subject of this paper.

- Management algorithms, including empty vehicle management and dynamic routing algorithms. In the present paper the former feature is addressed.

Many algorithms for empty vehicle management known from the literature are focused mainly on the optimal reallocation of empty vehicles, since this may reduce the passenger waiting time significantly (at the expense of increased empty vehicles movement). Reallocation algorithms are usually based on past demand estimates and future forecast [6-14]. All these approaches use a form of central repository in which historical demand and actual empty vehicles supply are stored. The demand forecasts are based on statistics from previous corresponding periods corrected for weather and special events. Due to the central data base, the algorithms are centralized. They mainly refer to empty vehicle reallocation (other cases are addressed separately) and hardly can be implemented in a distributed way.

Some papers [8,14] take into account other features, like the distance between stations or the time in which a vehicle may be delivered (in fact, this limits the distance as well). Yet, none gives a set of factors and thresholds to tune an algorithm to a given network with a given number of vehicles and a given demand. The work presented in [12,13] is extended in [15], where the traffic needed to move full and empty vehicles to their destinations is limited by the line capacities, taking into account speed limits and static/dynamic routing.

Our algorithm is based on the same general idea of demand forecast, as in [6-14], with a modification introduced in order to minimize the cost of empty vehicle management by limiting the distance of empty trips. As our algorithm does not use the complete information of positions of the vehicles (the decisions made with partial knowledge limited by the size of the horizon), we expect it will not perform optimally compared to other algorithms which use global knowledge. Yet, our algorithm gives significant improvements, as will be shown in section 5. The performance comparison with other algorithms is presented in concluding remarks.

The novelty of our algorithm lays in its easy-tuning due to many parameters being adjustable. For other algorithms, the parameters are hidden in the implementation. For example, when current demand does not fit past values, the "number of waiting passengers" may take the leading role in the reallocation of vehicles (sending additional vehicles to the stations which received additional demand).

Another example: a station with high demand monopolizes the addressing of empty vehicles, but a "number of empty berths" factor allows sending at least one vehicle to a less popular station. Simulations provide the values for the best setting of the parameters.

The repertoire of tasks needed for management algorithms is significantly larger than for reallocation only. Generally, empty vehicles management algorithms are used for:

- *calling* vehicles for new passengers if there are no empty vehicles presently available at a given station,

- *expelling* empty vehicles resting at the station in order to make room for approaching full vehicles when there are no free berths available to disembark the passengers at a given station,
- *balancing* the distribution of empty vehicles over the network in order to make them more easily accessible for subsequent use (reallocation),
- *withdrawing* empty vehicles to capacitors (e.g., for safety reasons, cleaning, maintenance purposes, etc.).

Similar division of empty vehicle management tasks into balancing and expelling is addressed in [16], although the algorithms themselves are not described there. In the present paper, a uniform algorithm used for all these tasks is discussed.

The role of balancing algorithm and its influence on both mean waiting time and the number of empty trips are commonly recognized and well known. However, the algorithm discussed in the present paper has some specific and advantageous properties. The most important are following:

- *parameters reflecting the state of the network* – in addition to the network topography and data on (current and past) transport demand, the algorithm takes advantage of the information on the state of stations, such as passenger queues lengths, the presence of empty vehicles and the availability of empty berths. The timing, origins, and destination of empty vehicle trips are determined by applying factors and thresholds to such indicators, i.e. the algorithm's parameters;
- *uniform procedure for many functionalities* – there is a generic procedure to schedule empty vehicle trips, and it is possible to implement different functionalities through parameterizing this generic procedure by specifying an individual set of factors and thresholds for every functionality. A distinction is made between different functionalities of the empty vehicle management (calling, expelling, balancing, withdrawing);
- *distribution* – the algorithm is distributed both in the terms of control and in the terms of data. The decisions for executing empty trips are made autonomously by the controllers of the stations (control distribution). The exchange of information is performed only among stations within a predefined range, called "horizon" (data distribution). This means that the management of empty vehicles requires neither a central controller nor a centralized collection of data;
- *local communication* – the algorithm is based on (assumed) communication of every station controller with its neighboring ones to receive their states, instead of communication with a central controller. Maximum "horizon", limiting the distance of communication, is defined as one of algorithm parameters. During the execution of the algorithm, traffic parameters' values are exchanged between stations (for example: the number of empty berths, the number of passenger groups waiting, the number of vehicles going to a station, etc.), but these data are sent by a station only to the stations not farther than the specified horizon;
- *economy* – the finite value of the horizon limits the total length of empty trips: limiting the distance of empty trips is an additional target of the algorithms [8,14], but in our algorithm this target is achieved automatically;
- *resilience* – the horizon may be tuned in a given PRT network and may be dynamically adjusted when the traffic conditions change (for instance, when average waiting time excessively grows; the manipulation with the horizon is described in [17]);

- *flexibility* – observation of various traffic parameters in addition to demand forecast – this allows for dynamic response to unusual traffic conditions, when demand forecast only complicates the operation of the system instead of facilitating it.

Numerous papers discuss pros and cons of centralized and distributed control, especially in transport systems [18,19]. In centralized systems, the control server is a single point of failure, but the effect of a crash is total immobilization. Distributed system has many points of failure, yet the controller crash has restricted impact on the functionality of the system. Centralized algorithms ale often more complicated and run slower, but a global maximum of the target function may be achieved. Conversely, distributed algorithms are faster and cheaper, but the global maximum may not be reached and they may sometimes lead to pathological situations. Distributed system acts in real time and is more reactive, while centralized one operates on a schedule basis. As all algorithms mentioned in the literature [6-14] are of centralized nature, our research aimed to prepare an algorithm suitable for the distributed implementation. Another attempt to build a distributed management algorithm is presented in [20], but in this paper the activity is attributed to the vehicles rather than to the stations.

The research on control algorithms for PRT networks was a part of Eco-Mobility project, done at the Warsaw University of Technology [21,22]. The methodology for testing and evaluating various versions of control algorithms has been based on discrete event simulation. For the purpose of this research we decided to build an own, proprietary PRT network simulator, named Feniks. The other available simulators [23-27] usually offer a fixed, inaccessible traffic algorithms so that the user can only observe their operation and there is no (or little) influence possible on algorithm structure and parameters. The fixed rules in those simulators include joining the traffic from a station, 'keeping up' algorithm, 'join' intersection priority rules, dynamic routing, vehicle calling rules, empty vehicle reallocation, expelling vehicles (or no expelling at all), etc. Additionally, the proprietary simulator provides a better access to individual events, which should be reported as a type of "event log", with event types defined by the researcher, allowing the researchers to build their own detailed characteristics of the traffic, etc.

The simplified version of the proposed multiparameter algorithm (without forecast) was described in [28]. The simulations were performed with a set of fixed parameters' values to show that even without using historic demand, the passenger waiting time may be reasonably shortened. Also, a procedure for obtaining maximum ridership of the network is presented there.

We tested the algorithm working in the network with extremely asymmetric demand (on individual stations) and Origin-Destination Matrix, while delivering participants to a social event on a suburbium and shipping them back home. Our research shows that the manipulation with the horizon value is enough to accommodate the network to such conditions [17].

To test various aspects of control and coordination in PRT system, some set of reference models (or benchmarks) is needed. Such a set is available, mainly of "geometric" nature, proposed by Lees-Miller [29]. We proposed our own set of benchmarks [30], which are rather "socially" motivated. The proposed distributed algorithm was tested on our benchmark models: *City*, *SeaShore* and *TwinCity* from [30] (detailed results for *City* model are presented in section 5) and on *Corby* and *Grid* models from [29] (for comparison with other algorithms, results in section 7).

## 1. PRT simulation environment

PRT network can be depicted as a graph consisting of multiple nodes and directed arcs, additionally described by a set of parameters. The simulator (Feniks) provides a graphical interface for edition of the graph (optionally – over the background map of the area), defining the parameters of network elements as well as setting the options and parameters of a simulation run.

The nodes in a PRT graph are of three types [3]: stations, capacitors and intersections. Capacitors and stations have one entry and one exit. They include parking places or berths. Stations are places where passengers order their trips and board onto the vehicles, or wait for a vehicle in a queue if there is no empty vehicle presently available. Passengers board on and alight from vehicles at berths. A station is characterized by the number of berths and by the sizes of entry buffer (where the incoming vehicles may wait for an empty berth) and exit buffer (where the outgoing vehicles wait if they cannot join the traffic due to traffic conditions).

A capacitor (or a garage) is an initial source of vehicles and sometimes may serve as a parking place. The number of parking berths is the only parameter of a capacitor.

Intersections are of two types: simple 'fork' (1=>2) and 'join'(2=>1).

It is assumed that each vehicle has its own control unit linked via radio network to the control units of other vehicles and nodes (stations, capacitors and 'join' intersections). This way, the vehicle gets information about the current values of the parameters of movement of preceding vehicles: their positions, velocities and mode of operation (acceleration/ constant velocity/ deceleration/ friction braking). The vehicle receives a decision on priority when crossing an intersection from the controller of this intersection. The issue of the priority on join-type junctions is not discussed in more detail because just the first, preliminary simulations shown that the priority rule applied (left, right, random or otherwise determined) is immaterial from the viewpoint of the purpose of the paper.

Nodes (capacitors, stations and intersections) are connected by unidirectional segments of a track system. There are two types of segments. Highway segments connect intersections only (no station is located immediately at the highway) and the velocity of vehicles on a highway is relatively high (typically maximum 15 m/s). Road segments may connect all types of nodes, and the velocity is lower (maximum 10 m/s). Maximum velocity may be additionally limited in some segments, depending on the terrain conditions. The parameters of a segment are: type (highway or road), length and maximum velocity.

Feniks is a discrete event simulator [31], based on microsimulation, therefore track segments are logically divided into equal-length sectors for the purpose of the simulation. The length of these sectors is defined at the simulator level. All the control decisions are made on sector connections. Of course, the sectors should be short enough to allow a simulation of relatively "smooth" traffic, at the expense of simulation run time [32]. In practice, it was assumed that the sector length was at least twice (or three times) smaller than safety distance (separation) between vehicles.

Each vehicle can accommodate up to 4 passengers. The vehicles perform trips between stations and/or capacitors. Trips are either full (i.e. with passengers) or empty. Although one can consider several types of vehicles (e.g. personal ones, for disabled persons, for maintenance purposes etc.), we assumed that all the vehicles are of one type, characterized by common parameters. The dynamic parameters of a vehicle are: maximum velocity on a highway and on a road, maximum acceleration and deceleration, minimum inductive deceleration and maximum friction deceleration (emergency brake), minimum separation between vehicles.

It is assumed that the passengers arrive at network stations in groups of 1 to 4 persons. Number of persons in a group is a random variable with a uniform distribution (i.e. mean is equal to 2.5). Each group performs a trip (i.e. travels together) to the common destination (no ride sharing). The input stream of passenger groups is random for the whole network as well as for individual stations, with exponential distribution of the inter-arrival time between groups. The mean of the distribution may differ between stations and may vary during the day.

Whenever a passenger group arrives at a station (in Fig. 1), it takes one of empty vehicles, if available. If there is no empty vehicle available, the passenger group waits in a queue (one at a station, common for all berths) until a vehicle becomes available, either because it concludes its trip at this station or is delivered by the empty vehicles management algorithm. Then, a group of passengers occupies the vehicle, makes a trip to the destination, and leaves the network. The boarding and alighting times are random variables with triangular distributions identical for all stations. The triangular distribution of the boarding and alighting time reflects the variation of group cardinality and other factors affecting the boarding and alighting.

Every passenger group chooses its own destination of the trip. In the simulator, the probability of undertaking a trip between every pair of stations is defined as an Origin-Destination Matrix (*ODM*). The *ODM* matrix may vary during the day.

The routing follows the Dijkstra's algorithm [33], with segment costs that are a combination of segment length, segment free passage time and traffic density (dynamic component). The three cost components are normalized (respectively: by average distance, by average free-path trip time between every pair of stations, and by number of vehicles in the model) and have user-defined weights. The simulator allows for dynamic routing to avoid traffic jams, but this feature will not be analyzed in the present paper.

In addition to the network graph and numerical model parameters (e.g., velocities, acceleration, separation, parameters of triangular distribution of boarding and alighting times, passenger input, *ODM* matrices etc.), Feniks supports the definition of various parameters of the simulation run (e.g., length of warm-up period, termination conditions, output data collected etc.). A more detailed description of Feniks simulator can be found in [21,22].

**2. Motivation and general rules of the algorithm**

The motivation for a new heuristic algorithm for empty vehicles management can be summarized as follows. The algorithm is used for several tasks:

- *Calling* – for delivering a vehicle to the station where a new group of passengers has just appeared while there is no empty vehicle available at this station,
- *Expelling* – for removing an empty vehicle from a station where there are no free berths while some vehicle approaches the station,
- *Balancing* – for moving the vehicles (in advance) to new stations where the *future* calls are expected,
- *Withdrawing* – for sending the vehicle back to capacitor (garage) for safety reasons, maintenance etc.

Regardless of which of above tasks has to be performed – the algorithm repeatedly performs just a few typical actions: selection of a particular vehicle, determination of the origin and the destination of a trip and initiation of the vehicle movement. From this moment on, the vehicle is controlled by other algorithms (dynamic routing, coordination etc.). Therefore, the main activity of empty vehicle management algorithm should follow the same basic procedure of planning empty trips.

On the other hand, for each of these four tasks the decisions are taken on the basis of different parameters. Therefore, we decided to collate the relevant parameters into four sets of similar, uniform structure, each referring to specific characteristics of a given task.

An important requirement was that, for the purpose of this research, the algorithm should allow to test its efficiency for various values of network state parameters affecting the need for reallocating the vehicles, such as passenger queue lengths or numbers of empty vehicles staying at individual stations. Therefore, a vector of weight factors and thresholds has been defined. Manipulation of these weights and thresholds allows the designer to make the decision-making procedure depending on different criteria.

We also required that the algorithm should not use the global information on the network state, actual position of all vehicles, lengths of all queues etc. Instead, it has to take decisions upon the limited, local information on the state of network objects laying within a user-defined "horizon" only. This would provide the algorithm its scalability.

Finally, notice that the importance and necessity of above-mentioned four tasks is not equal. *Calling* is obligatory and comes just from the nature of PRT. *Expelling* is not equally important, yet, without it, jams may occur at a station entry (see Fig. 1). When all of the berths and the whole entry buffer at a station are occupied, there is no room for an approaching vehicle inside the station and the vehicle must wait before entry. The other subtle situation refers to an approaching vehicle with passengers aboard: a room in entry buffer is not enough because passengers should not wait for the emptying the entry buffer. A room must be made for the approaching vehicle as well as for all empty vehicles preceding it in the entry buffer. *Balancing* is optional but it may shorten passenger waiting time if reallocated vehicles reach the stations where passengers would arrive and before normal calling procedure would supply vehicles. *Withdrawing* is an example of optional activities, for example for vehicle cleaning or preventing damage to vehicles staying at stations in underpopulated areas at night.

### 3. Algorithm description

Accordingly, as discussed above, four following functions were defined: 'calling function', 'expelling function', 'balancing function' and 'withdrawing function'. All these functions are based on a common scheme, and they differ only in some details. Parameters reflect some constant, static features of the model (for example, a total number of vehicles) as well as the dynamically changing, current state of the model (for example, a current number of passenger groups in a queue at a given station). The set of parameters that gives the shortest passenger waiting time among simulated sets for a given model, with a given demand and a number of vehicles, is called the *best* parameters' values set.

The following static and dynamic model parameters are used in each of the afore mentioned functions:

- $N$ – number of stations in the model,
- $NG$ – number of capacitors in the model,

- $J$ – number of vehicles in the model,
- $H_i$ – number of berths at station $s_i$ or capacitor $g_i$.
- $S=\{s_1..s_N\}$ – the set of stations,
- $G=\{g_{N+1}..g_{N+NG}\}$ – the set of capacitors,
- $V=\{v_1..v_J\}$ – the set of vehicles,
- $Q_i$ – current number of passenger groups in a queue at station $s_i$.
- $K_i$ – current number of vehicles in berths of station $s_i$ or capacitor $g_i$,
- $L_i$ – current number of empty vehicles in berths and in the entry buffer of station $s_i$ or capacitor $g_i$,
- $Z_i$ – current number of vehicles on a trip to station $s_i$ or capacitor $g_i$,
- $D_{ij}$ – [m] shortest distance from station $s_i$ to station $s_j$ or capacitor $g_j$,
- $D_{av}$ – [m] average distance between a pair of distinct stations,
- $ND_{ij}$ – normalized inverse distance between stations $s_i$ and $s_j$ (or capacitor $g_j$); $ND_{ij}=D_{av}/D_{ij}$. Notice that the distance between stations $D_{ij}$ is a *denominator*: the shorter the actual distance is - the greater is value of $ND_{ij}$; $ND=1$ for mean distance,
- $PI_i$ – [s] mean value of passenger groups inter-arrival time distribution at station $s_i$ at the same hour of day in previous days.

For the calculation of the values of the four functions, a vector of weighting factors and threshold values has been defined. The weighting factors determine how strongly given parameters influence the decision to move a given vehicle. The thresholds define minimum values of measured features which allow a vehicle to move.

- $F_Q$ – passenger queue factor – determines the impact of the passenger queue length in target station;
- $F_{EB}$ – empty berths factor – the impact of a number of empty berths in target station or capacitor;
- $F_{ND}$ – normalized inverse distance factor – the impact of normalized inverse distance between nodes (see $ND_{ij}$ above).
- $F_{AI}$ – historical input factor – the impact of mean value of passenger groups inter-arrival time distribution at target station during previous days (a measure of predicted demand); the mean value is a denominator (see the equation (1) later), because the shorter is the time between occurrences of two consecutive groups, the stronger the impact is;
- $T_Q$ – passenger queue threshold – if in a queue there are less passenger groups than $T_Q$, then a vehicle is not moved;
- $T_{EB}$ – empty berths threshold – if there are less empty berths than $T_{EB}$, then a vehicle is not moved;
- $T_{EV}$ – empty vehicles threshold – if there are less empty vehicles in berths than $T_{EV}$, then a vehicle is not moved;
- $T_{ND}$ – normalized inverse distance threshold (inverse of the horizon) – if the distance between nodes is greater than $T_{ND}$ (note that the actual distance between nodes is a denominator), then a vehicle is not moved;

- *T* – total function threshold – if the value calculated as the sum of products of individual factors by corresponding static or dynamic parameter values is less than *T*, then a vehicle is not moved.

Each function has its separate set of the above weighting factors and thresholds. The factors and thresholds for the balancing function have *B* prefix (i.e. $BF_Q$, $BF_{EB}$ etc.), for the calling function they have *C* prefix (i.e. $CF_Q$, $CF_{EB}$ etc.), *E* for the expelling function and *W* for the withdrawing function.

The evaluation rules are almost identical for each function. Below, evaluation rules of the balancing function are given. To obtain the analogous rules for the expelling function, one should only replace the prefix *B* by *C*, *E* or *W* in the appropriate remaining formulas.

As an example, let's take the balancing function (*B*) under consideration. The general idea is that the algorithm consists of the three following steps:

- Identify a subset of stations among which the balancing would be possible and beneficiary. The decisions are based on the actual queue lengths, the number of free berths, the distances between the stations etc., compared with appropriate thresholds.

- Within such a restricted subset, identify the station which is the best candidate for balancing. The evaluation is based on the value of special balancing function *B*, which is a weighted sum of station state indices and their weights $BF_Q$, $BF_{EB}$, etc.

- Perform an empty trip to the "best" station, provided that the just-computed "best" value of balancing function $B_{max}$ exceeds some general threshold *BT*.

More specifically, the procedure is as follows: for a given station $s_x$, we select a subset of stations $s_i$ (out of the set of all stations) for which the following conditions simultaneously hold:

- $Q_i - L_i - Z_i \geq BT_Q$ – queue length is greater than (or equal to) 'passenger queue threshold'; the number of empty vehicles at the station $s_i$ and the number of vehicles on a trip to the station $s_i$ are subtracted from the queue length, because these vehicles are about to take passengers soon,

- $(H_i - K_i + Q_i - Z_i)/H_i \geq BT_{EB}$ – normalized number of empty berths is greater than (or equal to) 'empty berths threshold'; the number of passenger groups in a queue is added to the number of free berths $H_i - K_i$: they will take vehicles soon; and the number of vehicles on a trip to the station $s_i$ are subtracted from the number of free berths $H_i-K_i$, as they will take empty berths,

- $ND_{xi} \geq BT_{ND}$ – normalized inverse distance is greater than (or equal to) 'normalized inverse distance threshold', i.e. stations closer than $1/BT_{ND}$ are considered,

- $(L_i + Z_i - Q_i)/H_i - (L_x + Z_x - Q_x)/H_x \geq BT_{EV}$ – the difference between fractions of empty vehicles in both stations is greater than (or equal to) 'empty vehicles threshold'.

Then, for every selected station $s_i$, the value of the balancing function ($B_i$) is calculated:

$$B_i = BF_Q*(Q_i - L_i - Z_i) + BF_{EB}*(H_i - K_i + Q_i - Z_i) + BF_{ND}*ND_{xi} + BF_{AI} / PI_i \qquad (1)$$

Note that the parameter $BT_{ND}$ plays a special role: it defines a distance to the stations that are taken into account in the algorithm (the horizon). All features other than *ND* (number of vehicles, number of free berths, number of passenger groups, etc.) can be obtained by means of communication of $s_x$ with other stations. No central database holding these data is needed. Therefore, in real world, the management algorithms can be implemented in a distributed way, basing on inter-station communication restricted to distance $1/BT_{ND}$.

A station $s_{max}$ with the highest value of $B_i$ (called $B_{max}$) is chosen as a candidate for effective balancing. Then, if $B_{max} \geq BT$, an empty trip of the vehicle from station $s_x$ to $s_{max}$ is executed. This decision is made for a randomly chosen vehicle at the station $s_x$. This vehicle is treated as *on the trip* to destination station $s_{max}$, therefore, the ongoing analysis for station $s_x$ finds one vehicle less staying at the station and one vehicle more traveling to destination station $s_{max}$. The procedure of the algorithm running on station $s_x$ is presented in pseudocode below:

```
int max =0;
float Bmax=-∞
for (i in 1..N)
{
        if (NDxi ≥ BTND)
        {
                acquire Qi , Li , Zi , Ki from station si;
                int targ_fun= BFQ*(Qi – Li – Zi) + BFEB*(Hi – Ki + Qi – Zi) + BFND*NDxi + BFAI / PIi ;
                if (Qi – Li – Zi ≥ BTQ &&
                   (Hi – Ki + Qi – Zi)/ Hi ≥ BTEB &&
                   NDxi ≥ BTND &&
                   (Li + Zi – Qi)/Hi – (Lx + Zx – Qx)/Hx ≥ BTEV &&
                   targ_fun  > Bmax)
                {
                        Bmax= targ_fun;
                        max=i;
                }
        }
}
if (max>0)  execute empty trip from sx  to smax;
```

The difference in the calling function (*C*) is that the empty trip is executed from station $s_{max}$ to $s_x$ rather than from station $s_x$ to $s_{max}$ as in (*B*). For withdrawing (*W*) – capacitors are considered instead of other stations. For expelling (*E*) – stations as well as capacitors are considered.

The described procedure is performed:

- for *balancing* – periodically,

- for *calling* – when a passenger group arrives and there is no empty vehicle, or when a vehicles finishes its trip (it is empty or becomes empty),

- for *withdrawing* – when an empty vehicle stays in a station longer than a specified timeout,

- for *expelling* – each time a vehicle approaches a station where all berths are occupied and at least one empty vehicle is at the station, this may not occur if all vehicles are during boarding or alighting – they cannot be moved.

## 4. Organization of simulation experiments

The algorithm has been verified during multiple simulations of various network models. However, in order to analyze the properties of the algorithm itself rather than estimate the performance indices of some specific network – we decided to perform a set of simulations of an artificial, benchmark model [30] called *City* (Fig. 2).

The topography of the model schematically represents a simplified layout of a traditional city, with its downtown (a smaller central zone) surrounded by eight suburbs or residential areas. Circles marked with letters represent 12 stations (A...D in central zone, E...L in suburbs) and dashed circles represent 4 capacitors. Double lines are two-way highways while single lines with arrows are one-way road segments. The diameter of the highway ring is 3 km. The total length of the track segments is about 33 km.

In addition to the network topography, several other properties were fixed for all experiments as well. They include the general model of demand and general rules of choosing the destination of full trips (*ODM*).

As it was mentioned in Section 1, the passengers arrive at the network in groups of 1 to 4 people. Each group travels together to the common destination. No ride sharing is allowed. Number of passengers in a group is random with a uniform distribution (thus the mean is equal to 2.5). Arrival (input) stream is the Poisson process, i.e. the inter-arrival time is an exponentially distributed random variable. The exponential rule is valid for input stream of the whole network as well as for individual stations, but the demand is not uniformly distributed over the network. Four stations receive input stream with an average rate computed for a single station (see below), the next four stations (randomly selected and then fixed for all simulations) have an input rate equal to 2/3 of this average, and the remaining four stations have 4/3 of average.

Also, the Origin-Destination Matrix (*ODM*) is not uniform. Every cell of the matrix (except for the main diagonal) was assigned a random value from range <0…1>, then the matrix was normalized to achieve a sum of 1 in every row. Sums of columns in the matrix range from 0.58 to 1.44, so that the "most popular" stations are selected as the destination almost 2.5 times more frequently than the "least popular" ones.

The above-mentioned properties apply to all simulations. However, during experiments several *variants* of the simulation model were used. Individual variants differ in the values of two following parameters:

- number of vehicles *J*,
- input rate λ (intensity of stream of passengers arriving to the network).

For each variant as much as 10 simulations were performed, differing in the configuration of four parameters of the balancing algorithm ($BF_{EB}$, $BF_Q$, $BF_{ND}$ and $BF_{AI}$). The organization of the whole set of simulation experiments will be discussed in more detail below.

The output (observed) data from each simulation run were:

- average squared passenger waiting time *ASWT*,
- number of empty trips *NET*.

We take waiting time squared because a simple average (*AWT*) does not properly reflect the level of passengers satisfaction. If average waiting time is acceptable, but a single passenger waits for an hour or longer, local newspaper or social media will surely inform about this fact with an unfavorable comment. The squared measure (in fact, root average squared, but we shortened the name for simplicity) measures passengers satisfaction better than a simple average ($T_{ft}$ is a waiting time before a given full trip):

$$ASWT = \sqrt{\frac{\Sigma_{full\,trips}\,T_{ft}^2}{number\,of\,full\,trips}} \qquad (2)$$

The general goal was to analyze how the different algorithm parameters affect *ASWT* and *NET* in different variants of the model (i.e. for different numbers of vehicles and different input rates). In particular, it was interesting if there exists a "best" configuration of balancing algorithm parameters that would provide a shorter *ASWT* with at most a minor, acceptable increase in *NET* . The numerical measure of the quality of the configuration *QC* is taken as the product of these two values:

$$QC = ASWT * NET \qquad (3)$$

The number of vehicles should be large enough to meet the demand with some safety margin for unusual situations (e.g., sudden increase of the input rate for a given station, etc.). On the other hand, too many vehicles increase the number of possible traffic conflicts resulting in jams or congestions. Some hints for choosing the right number *J* of vehicles for a given network are discussed in [28]. In accordance with these suggestions we decided to perform simulations for $J = 48$ and $J = 76$ vehicles.

In order to determine the range of variability of the input rate, two additional simulations were performed (for $J = 48$ and $J = 76$ vehicles). The goal was to estimate network's maximum ridership *M* [groups/h], i.e. the largest number of passenger groups that can be carried by the network. In these preliminary simulations (described in [28]), the input queue in every station was never empty so that each vehicle coming to a station was immediately boarded by a new group of passengers and undertook a new trip. This way no empty trip is executed and the algorithm for empty vehicles management is irrelevant. *ODM* matrix in these two experiments was uniform, i.e. passenger streams between every pair of stations were equal.

As a result**,** it was estimated that network's maximum ridership for 48 vehicles is $M = 638$ and for 76 vehicles $M = 986$. Obviously enough, *M* sets the absolute limit of network's equilibrium. Let λ [groups/h] be the (constant) assumed rate of random input stream for the whole network, and coefficient ρ be the ratio λ/*M*. As λ is growing – so are the waiting queues as well as average waiting time. The effect is especially dramatic for λ close to *M* (ρ close to 1) . For λ ≥*M* (ρ ≥1) the network is no longer in the state of equilibrium and the queues grow infinitely. Moreover, in practice:

- the demand is not uniform among the stations,
- the *ODM* matrix is not uniform,
- the structure of the network is usually not symmetric, resulting in an irregularity of the traffic on various segments,
- input rates and *ODM* matrix may change in time.

Therefore, in order to provide the safe, "normal" working conditions that would allow for analyzing properly the role of algorithm parameters – we assumed that in all variants of the model the input rate λ does not exceed 0.5*M*. This working range ρ [0, 0.5] was divided into several values so that in individual variants of the model the following network input rates were used [groups/h]:

- For 48 vehicles: 100, 155, 210 and 320 (ρ: 0.16, 0.24, 0.33, 0.50),
- For 76 vehicles: 150, 300 and 500 (ρ: 0.15, 0.30, 0.51).

The above values, divided by the number of stations, determine the average input rate *per station* in each variant. As it was mentioned above, in all 7 variants four stations receive the input stream with the rate equal to 2/3 of this average etc.

For each out of the 7 variants a set of 10 simulations has been performed. They differ in the use of four main parameters of the balancing algorithm, namely:

- empty berths factor $BF_{EB}$,
- passenger queue factor $BF_Q$,
- normalized inverse distance factor $BF_{ND}$,
- historical input factor $BF_{AI}$,

Other parameters were constant, namely:

- $BT_{EB} = 1/H_i$ (at least one empty berth),
- $BT_Q = -H_i+1$, where $H_i$ is the number of berths in the destination station (in order to not to take into account the passengers which are supposed to use empty vehicles staying in berths of the destination station, if any),
- $BT_{ND} = 1$ (execute empty trip only when distance is not greater than mean distance),
- $BT_{EV} = 0$ (neutral).

In consecutive simulations, four adjustable parameters (i.e. $BF_{EB}$, $BF_Q$, $BF_{ND}$ and $BF_{AI}$) were either "enabled" (i.e. set to their nominal values) or "disabled" (switched off) by resetting to 0. The set of 10 simulations (for each variant) included the following cases:

- all parameters off (reset to 0),
- all parameters on (enabled, i.e. set to 1, except for $BF_{AI}=5$ to highlight the importance of "prediction" represented by historical demand parameter),
- only one parameter on (the remaining parameters off, 4 experiments),
- only one parameter off (the remaining parameters on, 4 experiments).

For convenient reference, every experiment is identified below by a four-bit binary tag ($BF_{EB}$, $BF_Q$, $BF_{ND}$, $BF_{AI}$) where 0's correspond to disabled parameters while 1's – to enabled ones. For example (1101) is the tag of experiment in which parameters have the following values: $BF_{EB}=1$, $BF_Q=1$, $BF_{ND}=0$, $BF_{AI}=5$. Similarly, (0000) refers to the case when all four parameters are reset to 0, etc.

The total threshold $BT$ was set to 1, to guarantee that at least one of the changeable parameters "works" (multiplied by corresponding value in the model gives positive value).

The parameters of balancing function were the main topic of interest. Parameters of other management tasks were set to some "neutral" values which remained unchanged in all simulations. *Calling* parameters had the following values:

$$CT_{EB}=1/H_i, CT_Q= -H_i+1, CF_{EB}=0, CF_Q=0, CF_{ND}=5, CF_{AI}=0, CT=0$$

These values were chosen to deliver a vehicle if there is a group of passengers waiting (which will be not satisfied by vehicles traveling to the station) and there is an empty berth at the station. Shortest empty trips are preferred.

*Expelling* parameters were set to:

$$ET_{EB}=1/H_i,\ ET_Q=-H_i+1,\ EF_{EB}=1,\ EF_Q=1,\ EF_{ND}=1,\ EF_{AI}=0,\ ET=-\infty$$

This caused the empty vehicles to be expelled to the nearest station where there was a nonempty queue of waiting passengers and, simultaneously, there were empty berths while no vehicle was bound for them.

*ET* was set to a large negative value to allow expelling a vehicle regardless of situation, i.e. whenever the expelling was needed, it was executed (regardless of any parameter value).

*Withdrawing* was off (*WT* had very high value, so the values of other parameters were unimportant).

## 5. Simulation results

The results of the experiments are graphically illustrated in Figs. 3 and 4 (for 48 and 76 vehicles, respectively). Selected numerical values are collected in Tables 1 and 2.

Each plot corresponds to one *variant* of the model, i.e. it shows the results of a set of simulations with different algorithm parameters but the same number of vehicles and demand. Notice that such plots do not imply any causal relationship between *NET* (number of empty travels, horizontal axis) and *ASWT* (average squared waiting time, vertical axis). The small black diamonds indicate only the pairs of values of *NET* and *ASWT*, obtained experimentally from individual simulation runs.

In each plot, a point marked with a solid circle corresponds to a case when the balancing is completely switched off (tag 0000). For the purpose of Tables 1 and 2 (see below) this set of parameters is called $t_{base}$. Parameter values sets are scored using *QC* value. The best configuration is expected to be near lower left corner of the plot. A dashed circle highlights the configuration of parameters that provides the "best", i.e. the smallest *QC* ($t_{best}$). A dotted circle surrounds the point $t_{sug}$ which we define "suggested" tag because it is the best, or near the best, in all experiments for given variant. In other words, the "suggested" configuration of parameters fits to changing demand and traffic conditions, although it may result in a bit worse *QC* than the "best" one. This $t_{sug}$ (its tag is 1111) is the recommended configuration of the algorithm parameters.

The numerical values of *ASWT* and *NET* for $t_{base}$, $t_{sug}$ and $t_{best}$ are collected in Table 1 (for 48 vehicles) and Table 2 (for 76 vehicles). For $t_{sug}$ and $t_{best}$ the improvement compared to $t_{base}$ is shown ((($t_{base} - t_{best}$) / $t_{base}$)*100%; analogously for $t_{sug}$) Also, for $t_{best}$ the binary tag is given.

The experiment with demand λ>*M*/2 (ρ>0.5) is especially noteworthy. In this case, no configuration of parameters gives better results than $t_{base}$ (i.e. the case when balancing is switched off). This confirms the assumption that the demand should not exceed a half of network's ridership (*M*) for a given number of vehicles.

The discussion above suggests the procedure of dealing with algorithm parameters when a new case of the network is to be analyzed. Now, for a given network topography and the given number of vehicles (i.e. the variant of the network) the choice of proper configuration of parameters' values should be preceded by a set of simulations for various configuration tags and for several demand models (e.g. different arrival rates). The configuration finally selected as "suggested" would determine the strategy of empty vehicles management in the given network. "Suggested" strategy should represent, for instance, a balance between passengers' mean waiting time and the number and/or distance of empty trips. Generally, other criteria can also be considered. The choice of parameters' values may be performed simply by a set of simulations with specified target features.

Of course, one can prepare several (predefined) strategies and use of them depending on specific circumstances (time of day, special events etc.).

## 6. Sensitivity for parameters change and further experiments

The important question is how much the algorithm is sensitive to the changes to its parameters. To get the initial insight, a new set of 32 simulation runs were performed, organized as follows. First, the parameter vector $t_{sug}$ was selected: $BF_{EB}=1$, $BF_Q=1$, $BF_{ND}=1$, $BF_{AI}=5$. Then, simulations were performed with single parameter value changed by ±30%. The simulations were run for the following variants:

- $J$=48 vehicles, λ=100 groups/h,
- $J$=48 vehicles, λ=320 groups/h,
- $J$=76 vehicles, λ=150 groups/h,
- $J$=76 vehicles, λ=300 groups/h.

We do not give the detailed results here. Generally, for greater demand (320,300) the maximum growth for *ASWT* is 17.3% and for *NET* is 9.8%. For lower demand (100,150) the maximum growth for *ASWT* is 97.6% (yet it is still about three times better than without balancing) and for *NET* is 40.4%. Therefore, we conclude that the sensitivity of the algorithm to the changes in its parameters values is acceptable.

We tested the algorithm against symmetric and more asymmetric passenger flows. For this purpose, three series of simulations were prepared: one with uniform *ODM* (*ODM1*), the second with zeroed odd columns in *ODM* (experiment *ODM2*), and the third with every fourth column left non-zero only (experiment *ODM4*). Again, vector $t_{sug}$ was used. The results are collected in Table 3. For *ODM1*, the *ASWT* result is better, which is natural (improvement to "no balancing" is between 52.8% and 98.0 %). For *ODM2* and *ODM4*, the *ASWT* result increased by several hundred percent, but the improvement compared to "no balancing" (tag 0000) was still very high (from 29.2% to 97.8%). This shows that the algorithm worked well even if the *ODM* is unbalanced.

Every variant was checked for configuration $t_{sug}$ (tag 1111) with the 1/2, 3/2, and infinite horizon (every station communicates with every other one, long empty trips are possible). For the infinite horizon, *ASWT* increases significantly for some demands (from 6.8% to 80.3%), but the mileage of empty trips (*ETM*) does not grow much (from 0.2% to 24.8%) . Tests for horizon 1/2 and 3/2 show that the optimal value of the horizon is between these two values (for some demands closer to one of these values than to 1). These results suggest that the assumption to take the horizon =1 (limiting empty trip distance to average inter-station distance) is proper. The results are collected in Table 4. Double bold line shows the place of horizon=1 between 1/2 and 3/2 (the values for 1 are all 0%). The better performance for finite horizon may be explained by the fact that a smaller horizon prefers shorter empty trips and vehicles become available again quicker. On the other hand, a small horizon may cause empty trips to be deferred because no empty vehicles are present in a distance less than the horizon, and the situation may change in the meanwhile.

Similar experiments allow to obtain a minimum number of vehicles to serve a given demand. These experiments involve varying the number of vehicles, while keeping the demand constant. The input values of 100, 200, 300, 400 and 500 groups/h were tested, with number of vehicles between 20 and 80 and step 2. The minimum number of vehicles is taken from the simulation in which *ASWT* is lower than 300s (5 minutes of waiting – arbitrarily chosen value). The results are collected in Table 5. For every demand value, the value of

*ASWT* and the number of vehicles are shown below (column "enough vehicles") and above threshold 300s (column "too little vehicles").

However, the demand should not exceed ρ=0.5, and all numbers of vehicles supporting given demands are above the values of ρ allowed (column "enough vehicles"). Therefore, vehicle numbers from the "safe" column, with ρ<0.5, should be used.

## 7. Conclusions and further work

In our research, we tested the empty vehicles management algorithm based on the multi-parameter analysis of a current state of a PRT network. The predicted demand is included as one of many algorithm parameters, and its weight may be controlled by the value of an appropriate parameter. The results of our simulations show that the algorithm may cause a substantial shortening of passenger waiting time. Of course, the best results are obtained by the cooperation of demand forecast with the observation of current state of the network, but the simulations described in [28] show that even when the demand forecast is not used ($BF_{AI}$=0), *ASWT* may be shortened significantly. The lack of central repository of current requests and empty vehicle supply makes our algorithm feasible for large networks with distributed control. In contrast to a global data base, our multi-parameter algorithm uses inter-station exchange of data, with possibility of limiting the distance among communicating stations.

In the algorithm, we included weights and thresholds for all the features that may influence passenger waiting time: distance, number of passengers waiting, number of vehicles staying in berths or traveling to given station, number of empty berths and demand forecast. Other parameters may be added, provided that they can be somehow reported to the stations using inter-station communication (rather than collected in central data base).

The results show that the use of the described heuristic algorithm is advantageous. The important features of the presented algorithm are:

- For balancing (reallocation) it gives good results, even better that in the case of prediction only. This feature allows for empty vehicles reallocation even in a case of untypical behavior of people, for example before and after a rally, a community meeting, mass events etc. This case with a highly unsymmetrical model of demand has been investigated in another paper [16].

- It is not based on any centrally collected data. Instead, it makes the use of information interchanged between stations and vehicles, while one of the parameters allows for restriction of the maximum distance for which the exchange of information takes place. This approach seems to be well suited to the distributed implementation of control and supports scalability.

- It limits the load of communication in the case of large scale networks and restricts the effect of node crash or communication link crash.

- It supports the economy of empty vehicle reallocation: avoids excessively long empty vehicle trips.

- It guarantees that stations remain operational and that it is very likely that users find empty vehicles at all stations.

Moreover, the same, uniform idea of the algorithm is applied to other aspects of vehicle management (calling, expelling, withdrawing). The procedure is almost identical while the sets of parameters are specific. The meaning and purpose of the parameters are analogous between the tasks. Indeed, a single procedure is used in the implementation of the algorithm.

The disadvantages of the described algorithm are typical for distributed control, mainly that global maximum (minimizing waiting times globally) may not be reached. Performing optimization on the basis of global knowledge may give better results (see later, the comparison with other algorithms using global data base).

The complexity of the algorithm is not troublesome: every station or capacitor obtains relevant data (passenger queue length, number of empty berths, number of empty vehicles) from neighboring stations/capacitors located no more than $BT_{ND}$ (or, respectively, $CT_{ND}$, $ET_{ND}$, $WT_{ND}$), calculates the value of function $B$ (or $C$, $E$, $W$) for every node and finally compares these values to obtain $B_{max}$. Nodes perform this procedure in specific points in time (periodically for $B$, on user demand for $C$, after specified events for $E$ and after timeout for $W$) independently of one another.

The comparison of the efficiency of reallocation algorithms should be based on the numerical results of the simulations of *the same* network model. This suggests that, for the purpose of research on control algorithms, it would be very profitable to establish some set of reference models (or benchmarks) which could serve the research community as a testbed for comparisons and evaluation. The proposition of such a set of "social" benchmark models (*City* is one of them) is presented in [30]. The algorithm presented in this paper was tested on other models from the set of benchmarks: *TwinCity* and *SeaShore*. These benchmarks have different characteristics each (topography and size), and the algorithm works properly in both of them.

Lees-Miller presented a set of benchmarks ([29] , mainly of "geometrical" nature), used in the evaluation of four algorithms: NN, SNN, SD and SV. They were tested on two benchmark models: *Corby* and *Grid* in [12].

- NN – older algorithm Nearest Neighbors [12,34] – based on trip schedule dynamically being prepared for every vehicle.

- SNN – Static Near Neighbor [12] – similar to NN, but requests are known in advance.

- SD – Surplus/Deficit [12] – based on observation of cumulative average of all previous empty vehicle trip times to a station.

- SV – Sampling and Voting [12] – analyzing possible sequences of future requests generated from the demand matrix and voting between them).

We compared our algorithm (marked *distr*) with the above four algorithms on *Corby* and *Grid* networks. The results for *Corby* network are presented in Fig. 5. The upper plot is linear, and the lower one is a semi-logarithmic plot (with linear demand axis). Fig. 6 presents the results for *Grid* network. Because of the lack of squared *AWT* in the results of experiments with the NN, SNN, SD and SV algorithms [12], the non-squared *AWT* values are presented both for these algorithms and for our *distr* algorithm. For both benchmarks (*Corby* and *Grid*) SNN gives best results, because the requests are known in advance in this case. SV is the second. Our distributed algorithm is worse than SV (which collects global database of requests and vehicles) and comparable to SD (better for demand ρ <0.35, worse for higher demand). NN is always worst (outside the chart on linear plot, visible on semi-logarithmic plot only). For comparison, the results of our distributed algorithm with forecast only are presented (tag 0010 – the factors concerning the current state of the network are switched off). The results for (0010) are significantly worse than for tag (1111). Fig. 6 presents the results for the *Grid* model. They are similar, except for demand over ρ=0.55 when our algorithm is better than SD (but this demand is above the safety threshold ρ=0.5).

Finally we may note that the Feniks simulator (current version: 4.0) proved to be a very useful tool for the analysis of the behavior of vehicles in a PRT network and testing of management algorithms. Future experiments are planned with other aspects of vehicle management, dynamic routing, fault-tolerant properties and optimization of network operation.

In the future work, it will be useful to compare our method used for estimating maximum ridership and minimum number of vehicles with the linear programming and maximum flow methods in the literature (for example [15,34]). Similar multiparameter approach may be used in other aspects of PRT control like static and dynamic routing (as in [15]), intersection precedence rules, joining the traffic from the stations, etc. Also, the cooperation of PRT with other transport means, basing on timetable rather than on demand will be investigated. This requires further development of the Feniks simulator.

## 8. Acknowledgments

The research described in this article was partially carried out (namely, the Feniks simulator) under the project ECO-mobility, co-funded by the European Regional Development Fund within the framework of Innovative Economy Operational Programme (WND-POIG.01.03.01-14-154/09).

The *City* model topography and parameters are uploaded to Wiktor Daszczuk's page on ResearchGate, as supplementary material to [30].

**List of symbols and abbreviations**

| | |
|---|---|
| *ASWT* | average squared waiting time |
| *AWT* | average waiting time |
| $B_i$ | balancing function for station $i$ |
| $B_{max}$ | maximum value of $B_i$ |
| $BF_Q$, $BF_{EB}$, $BF_{ND}$, $BF_{AI}$, $BT_Q$, $BT_{EB}$, $BT_{EV}$, $BT_{ND}$, $BT$ – parameters for *balancing* function | |
| $C_i$ | calling function for station $i$ |
| $C_{max}$ | maximum value of $C_i$ |
| $CF_Q$, $CF_{EB}$, $CF_{ND}$, $CF_{AI}$, $CT_Q$, $CT_{EB}$, $CT_{EV}$, $CT_{ND}$, $CT$ – parameters for *calling* function | |
| $C_{max}$, $B_{max}$, $E_{max}$, $W_{max}$ | highest value of function $C$, $B$, $E$ or $W$ |
| $D_{av}$ | (meters) average distance between pairs of distinct stations |
| $D_{ij}$ | (meters) shortest distance from station $s_i$ to station $s_j$ or capacitor $g_i$ |
| distr | distributed algorithm described in the paper |
| $E_i$ | expelling function for station $i$ |
| $E_{max}$ | maximum value of $E_i$ |
| $EF_Q$, $EF_{EB}$, $EF_{ND}$, $EF_{AI}$, $ET_Q$, $ET_{EB}$, $ET_{EV}$, $ET_{ND}$, $ET$ – parameters for *expelling* function | |
| $F_{AI}$ | historical input factor |
| $F_{EB}$ | empty berths factor |
| $F_{ND}$ | normalized inverse distance factor |
| $F_Q$ | passenger queue factor |
| $G=\{g_{N+1}..g_{N+NG}\}$ | the set of capacitors |
| $g_i$, $g_j$, $g_x$ | capacitor |
| $H_i$ | number of berths at station $s_i$ or capacitor $g_i$ |
| $J$ | number of vehicles in the model |
| $K_i$ | current number of vehicles in berths of station $s_i$ or capacitor $g_i$ |
| $L_i$ | current number of empty vehicles in berths and in the entry buffer of station $s_i$ or capacitor $g_i$ |
| $M$ | (groups/h) maximum ridership of the variant |
| $N$ | number of stations in the model |
| *NET* | number of empty trips |
| *NG* | number of capacitors (garages) in the model |
| $ND_{ij}$ | normalized inverse distance between stations $s_i$ and $s_j$ (or capacitor $g_j$) |
| *NN* | Nearest Neighbors algorithm |
| *ODM* | origin-destination matrix |
| *ODM1* | symmetric *ODM* |
| *ODM2* | original *ODM* (random) with zeroed odd columns |
| *ODM4* | original *ODM* (random) with every fourth column left non-zero only |
| $PI_i$ | (seconds) mean value of passenger groups inter-arrival time distribution at station $s_i$ at the same hour of day in previous days |
| *PRT* | personal rapid transit |
| *QC* | quality of the configuration (*QC=ASWT\*NET*) |
| $Q_i$ | current number of passenger groups in a queue at station $s_i$ |
| $S=\{s_1..s_N\}$ | the set of stations |
| *SD* | Surplus/Deficit algorithm |
| $s_i, s_j, s_x$ | station |
| $s_{max}$ | station with highest value of a function (*C, B, E* or *W*); |
| *SNN* | Static Near Neighbor algorithm |
| *SV* | Sampling and Voting algorithm |
| $T$ | total function threshold |
| $T_{ft}$ | (seconds) waiting time before given trip |
| $T_{EB}$ | empty berths threshold |
| $T_{EV}$ | empty vehicles threshold |
| $T_{ND}$ | normalized inverse distance threshold |
| $T_Q$ | passenger queue threshold |
| $t_{base}$ | basic values of parameters |
| $t_{best}$ | values of parameters giving best results (best *ASWT*) |
| $t_{sug}$ | suggested set of parameters – best or nearly best in all simulations |
| $V=\{v_1..v_J\}$ | the set of vehicles |
| $W_i$ | withdrawing function for station $i$ |
| $W_{max}$ | maximum value of $W_i$ |
| $WF_Q, WF_{EB}, WF_{ND}, WF_{AI}, WT_Q, WT_{EB}, WT_{EV}, WT_{ND}, WT$ | parameters for *withdrawing* function |
| $Z_i$ | current number of vehicles on a trip to station $s_i$ or capacitor $g_i$ |
| $\lambda$ | (groups/h) – intensity of input stream of the variant (input rate) |
| $\rho$ | relative demand of the variant ($\rho=\lambda/M$) |

Table 1. Results of experiments with 48 vehicles

| λ [groups/h] | ρ | variable | RESULTS | | | | | |
|---|---|---|---|---|---|---|---|---|
| | | | $t_{base}$ 0000 | $t_{sug}$ 1111 | | $t_{best}$ | | |
| | | | value | value | improvement [%] | value | improvement [%] | tag |
| 100 | 0.16 | ASWT [s] | 75.0 | 15.9 | 78.9 | 21.2 | 71.8 | 0111 |
| | | NET | 357 | 735 | -105.9 | 473 | -32.5 | |
| | | ETM [km] | 61.7 | 99.6 | -62.2 | 70.2 | -13.7 | |
| 155 | 0.24 | ASWT [s] | 85.4 | 25.3 | 70.3 | 26.9 | 68.6 | 0111 |
| | | NET | 646 | 1138 | -76.2 | 745 | -15.3 | |
| | | ETM [km] | 115.0 | 146.9 | -27.7 | 111.4 | 3.1 | |
| 210 | 0.33 | ASWT [s] | 96.9 | 36.3 | 62.6 | 44.2 | 54.4 | 1101 |
| | | NET | 784 | 1500 | -91.3 | 860 | -9.7 | |
| | | ETM [km] | 143.5 | 205.8 | -43.4 | 159.8 | -11.4 | |
| 320 | 0.50 | ASWT [s] | 139.8 | 56.6 | 59.5 | 71.6 | 48.9 | 1101 |
| | | NET | 1201 | 2002 | -66.7 | 1333 | -11.0 | |
| | | ETM [km] | 220.4 | 301.2 | -36.7 | 250.7 | -13.7 | |

Table 2. Results of experiments with 76 vehicles

| λ [groups/h] | ρ | variable | RESULTS | | | | | |
|---|---|---|---|---|---|---|---|---|
| | | | $t_{base}$ 0000 | $t_{sug}$ 1111 | | $t_{best}$ | | |
| | | | value | value | improvement [%] | value | improvement [%] | tag |
| 150 | 0.15 | ASWT [s] | 83.6 | 15.6 | 81.3 | 15.6 | 81.3 | 1111 |
| | | NET | 558 | 874 | -56.6 | 874 | -56.6 | |
| | | ETM [km] | 98.4 | 118.0 | -19.9 | 118.0 | -19.9 | |
| 300 | 0.30 | ASWT [s] | 100.5 | 31.5 | 68.6 | 31.5 | 68.6 | 1111 |
| | | NET | 993 | 1456 | -46.6 | 1456 | -46.6 | |
| | | ETM [km] | 181.4 | 218.6 | -20.5 | 218.6 | -20.5 | |
| 500 | 0.51 | ASWT [s] | 134.6 | 49.3 | 63.3 | 49.3 | 63.3 | 1111 |
| | | NET | 1567 | 2078 | -32.6 | 2078 | -32.6 | |
| | | ETM [km] | 289.5 | 330.0 | -14.0 | 330.0 | -14.0 | |

Table 3. Experiments with symmetric and asymmetric ODM

| number of vehicles | λ [groups/h] | EXPERIMENT (% growth to original ODM and horizon=1) | | | | | |
|---|---|---|---|---|---|---|---|
| | | ODM1 | | ODM2 | | ODM4 | |
| | | ASWT [% growth] | improvement to 0000 [%] | ASWT [% growth] | improvement to 0000 [%] | ASWT [% growth] | improvement to 0000 [%] |
| 48 | 100 | -93.6 | 98.0 | 101.3 | 72.9 | 247.5 | 64.3 |
| | 155 | -10.5 | 68.0 | 72.5 | 67.3 | 124.2 | 88.4 |
| | 210 | -24.4 | 71.0 | 50.5 | 87.2 | 97.8 | 97.8 |
| | 320 | -2.6 | 52.8 | 96.4 | 93.3 | 798.3 | 83.3 |
| 76 | 150 | -68.7 | 92.8 | 147.8 | 69.4 | 195.3 | 83.2 |
| | 300 | -32.5 | 75.3 | 159.3 | 96.1 | 381.0 | 96.7 |
| | 500 | -14.7 | 58.4 | 427.4 | 59.0 | 607.5 | 29.2 |

Table 4. Experiments with various horizon values

| number of vehicles | λ [groups/h] | EXPERIMENT (% growth to original ODM and horizon=1) | | | | | |
|---|---|---|---|---|---|---|---|
| | | horizon = 1/2 | | horizon = 3/2 | | infinite horizon | |
| | | ETM | ASWT | ETM | ASWT | ETM | ASWT |
| 48 | 100 | 53.4 | 17.4 | 28.9 | -76.7 | 24.8 | 80.3 |
| | 155 | 9.7 | 150.1 | 47.6 | 16.7 | 21.3 | 33.2 |
| | 210 | -10.2 | 92.6 | 36.7 | 0.0 | 15.7 | 13.6 |
| | 320 | -10.5 | 73.4 | 38.5 | 10.3 | 7.1 | 7.9 |
| 76 | 150 | 80.1 | 96.6 | 100.4 | -51.2 | 0.2 | 19.6 |
| | 300 | 33.2 | 99.6 | 50.5 | -30.4 | 8.9 | 9.8 |
| | 500 | 22.1 | 66.5 | 42.0 | -17.2 | 3.6 | 6.8 |

**Table 5. Obtaining a minimum number of vehicles for a given demand, *ASWT* threshold 240s**

| λ [groups/h] | EXPERIMENT | | | | | | | |
|---|---|---|---|---|---|---|---|---|
| | below *ASWT*=300s (enough vehicles) | | | above *ASWT*=300s (too little vehicles) | | Safe ($\rho \leq 0.5$) | | |
| | Number of vehicles | ρ | *ASWT* [s] | Number of vehicles | *ASWT* [s] | Number of vehicles | ρ | *ASWT* [s] |
| 100 | 14 | 0.56 | 182.6 | 12 | 369.7 | 16 | 0.49 | 121.4 |
| 200 | 22 | 0.62 | 255.1 | 20 | 1210.9 | 32 | 0.49 | 68.8 |
| 300 | 32 | 0.74 | 180.4 | 30 | 421.8 | 48 | 0.49 | 50.7 |
| 400 | 42 | 0.74 | 182.7 | 40 | 677.2 | 64 | 0.49 | 40.9 |
| 500 | 50 | 0.78 | 227.9 | 48 | 728.0 | 80 | 0.49 | 36.4 |

**Figure 1:** The PRT station

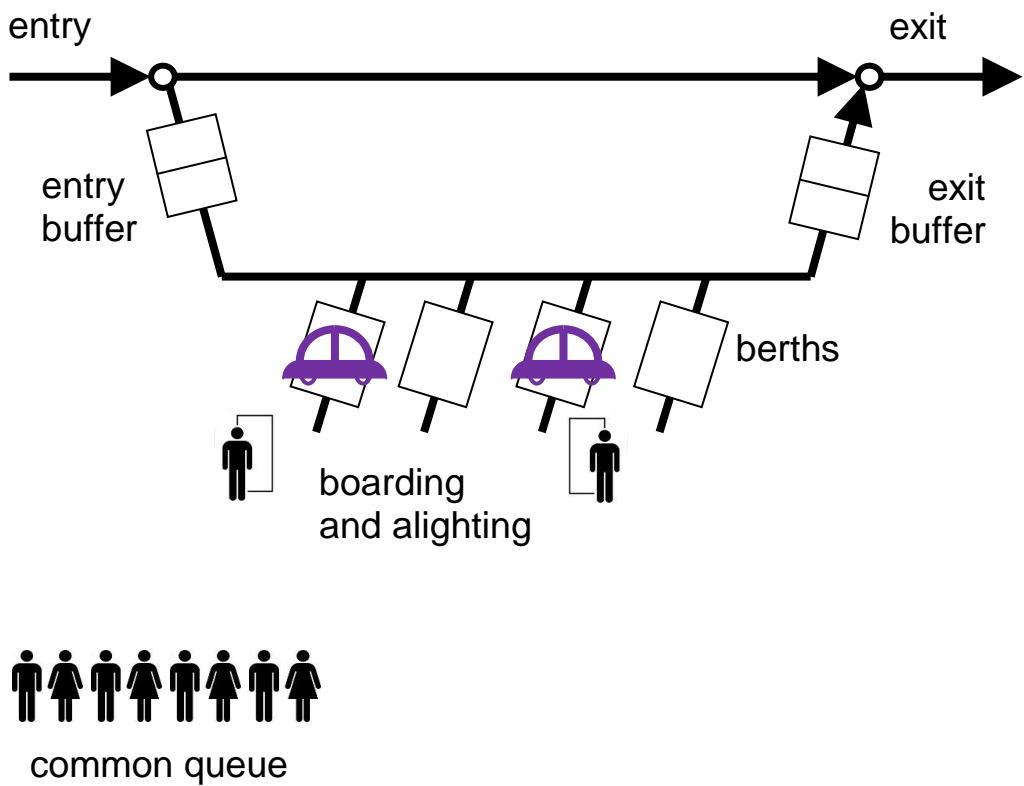

**Figure 2:** The "City" model

**Figure 3:** Average Squared Waiting Time (*ASWT* (seconds), vertical axis) versus number of Empty Trips (*NET*, horizontal axis) for various demand values (absolute λ and relative ρ), *J*=48 vehicles. Solid circles highlight the "base" configuration of parameters (tagged 0000), dashed ones – the cases providing the "best" (i.e., the smallest *ASWT*NET*), dotted circles – the "suggested" configuration of parameters (tag 1111) – always near the "best" for 48 vehicles.

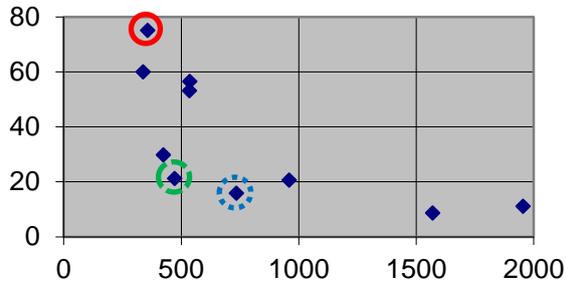

A. λ=100, ρ=0.16

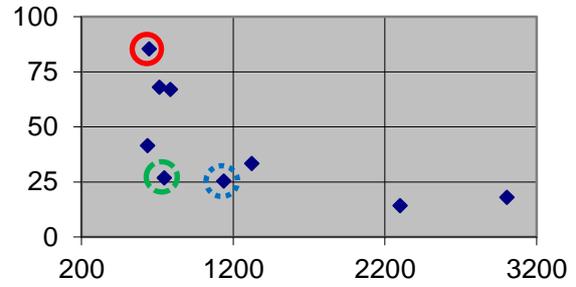

B. λ=155, ρ=0.24

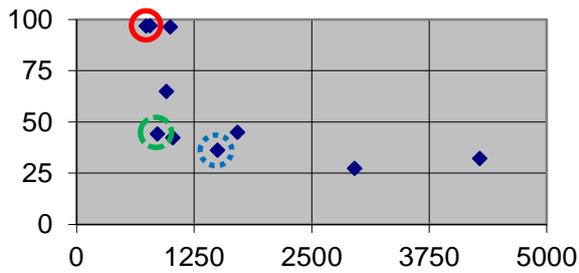

C. λ=210, ρ=0.33

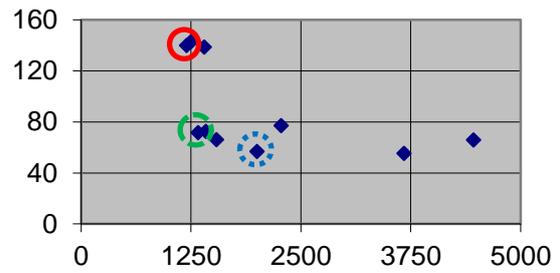

D. λ=320, ρ=0.50

**Figure 4:** Average Squared Waiting Time (*ASWT* (seconds), vertical axis) versus number of Empty Trips (*NET*, horizontal axis) for various demand values (absolute λ and relative ρ), *J*=76 vehicles. Solid circles highlight the "base" configuration of parameters (tagged 0000), dashed ones – the cases providing the "best" configuration (i.e. the smallest *ASWT\*NET*), dotted circles – the "suggested" configuration of parameters (tag 1111) – the same as the "best" for 76 vehicles.

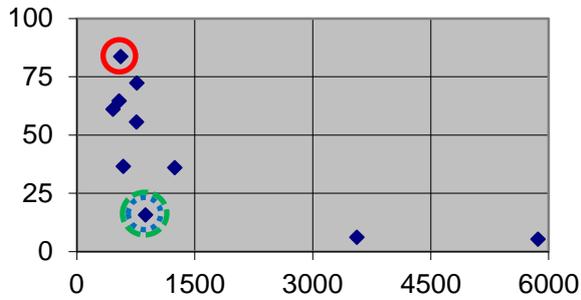

A. λ=150, ρ=0.15

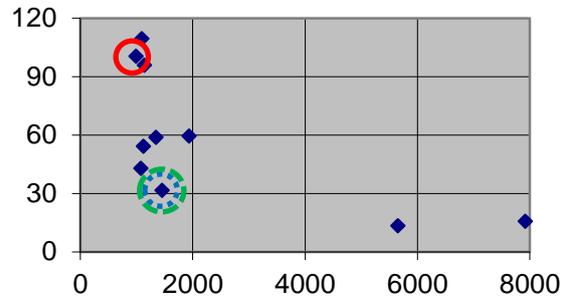

B. λ=300, ρ=0.30

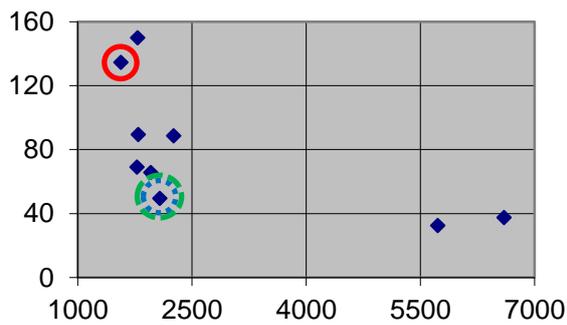

C. λ=500, ρ=0.51

**Figure 5:** Comparison of the effectiveness of the algorithms: distributed, distributed with forecast only, NN, SD, SNN and SV: linear plot (upper) and semi-logarithmic (lower) with linear ρ axis; *Corby* benchmark

**Figure 6:** Comparison of the effectiveness of the algorithms: distributed, distributed with forecast only, NN, SD, SNN and SV: linear plot (upper) and semi-logarithmic (lower) with linear ρ axis; *Grid* benchmark

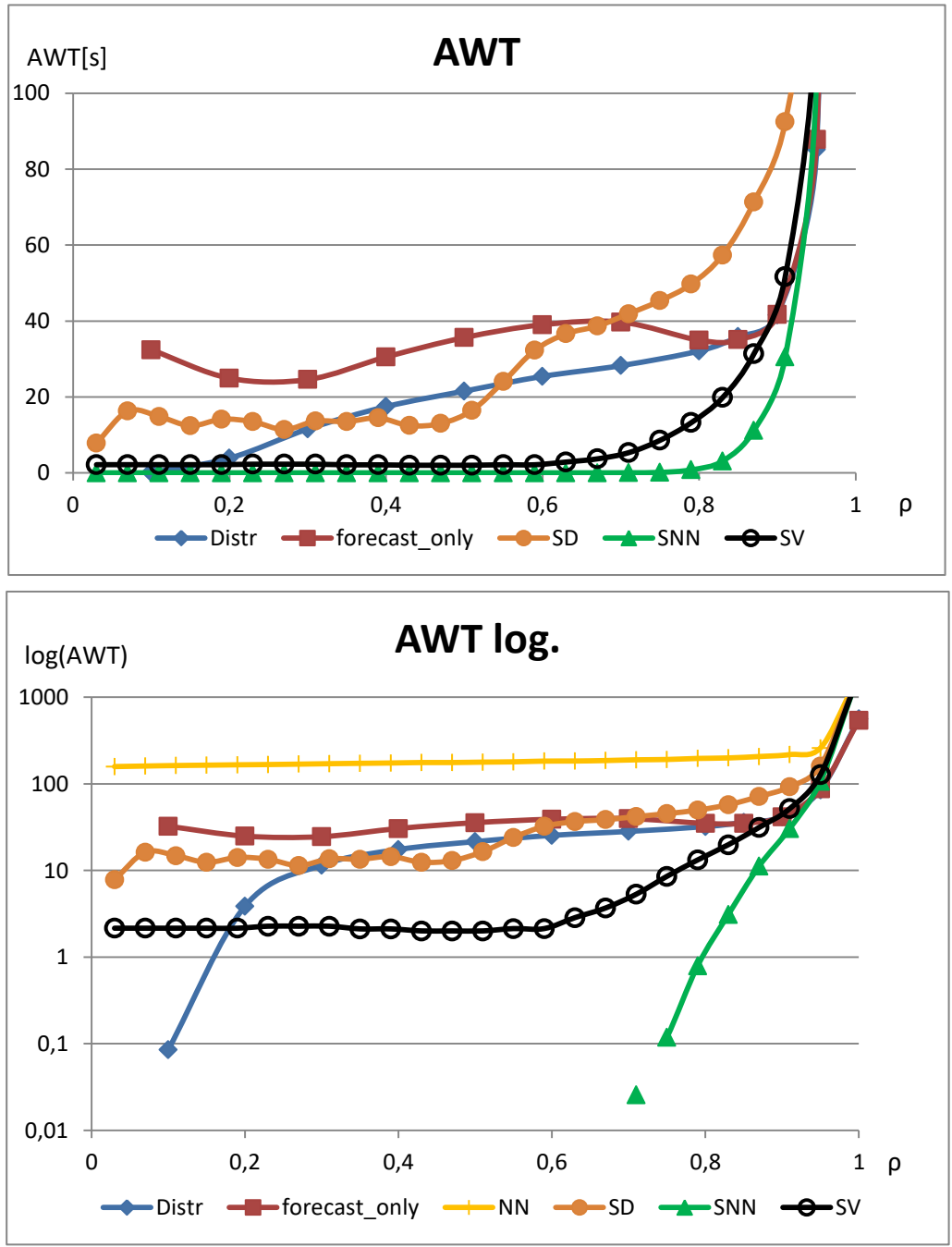